\documentclass{article}

\pdfoutput=1
\usepackage{arxiv}

\usepackage[utf8]{inputenc} % allow utf-8 input
\usepackage[T1]{fontenc}    % use 8-bit T1 fonts
\usepackage{hyperref}       % hyperlinks
\usepackage{url}            % simple URL typesetting
\usepackage{booktabs}       % professional-quality tables
\usepackage{amsfonts}       % blackboard math symbols
\usepackage{nicefrac}       % compact symbols for 1/2, etc.
\usepackage{microtype}      % microtypography
\usepackage{lipsum}		% Can be removed after putting your text content
\usepackage{graphicx}
\usepackage{natbib}
\usepackage{doi}

\usepackage[utf8]{inputenc} % allow utf-8 input
\usepackage[T1]{fontenc}    % use 8-bit T1 fonts
\usepackage{hyperref}       % hyperlinks
\usepackage{url}            % simple URL typesetting
\usepackage{booktabs}       % professional-quality tables
\usepackage{amsfonts}       % blackboard math symbols
\usepackage{nicefrac}       % compact symbols for 1/2, etc.
\usepackage{microtype}      % microtypography
\usepackage{xcolor}         % colors
\usepackage{amsmath}
\usepackage{cleveref}
\usepackage{graphicx}
\usepackage{multirow}

\usepackage{graphicx}
\usepackage{bbding}
\usepackage{algorithm}
\usepackage{algorithmic}
\usepackage{multicol}

\usepackage{wrapfig} 

\usepackage{xcolor}
\usepackage{colortbl}

\title{MODS: Multi-source Observations Conditional Diffusion Model for Meteorological State Downscaling}

\author{Siwei Tu$^{1}$, Jingyi Xu$^{1}$, Weidong Yang$^{1, }$\Envelope, \textbf{Lei Bai$^{2 }$}, \textbf{Ben Fei$^{3, }$\Envelope}\\
	$^1$Fudan University, $^2$Shanghai AI Laboratory, $^3$Chinese University of Hong Kong\\
    \texttt{tusiwei906@gmail.com},
    \texttt{wdyang@fudan.edu.cn},
    \texttt{benfei@cuhk.edu.hk}\\ \Envelope Corresponding Authors
}

\date{}

\renewcommand{\shorttitle}{MODS}

\hypersetup{
pdftitle={A template for the arxiv style},
pdfsubject={q-bio.NC, q-bio.QM},
pdfauthor={David S.~Hippocampus, Elias D.~Striatum},
pdfkeywords={First keyword, Second keyword, More},
}

\begin{document}
\maketitle

\begin{abstract}
	% \lipsum[1]
    Accurate acquisition of high-resolution surface meteorological conditions is critical for forecasting and simulating meteorological variables.
Directly applying spatial interpolation methods to derive meteorological values at specific locations from low-resolution grid fields often yields results that deviate significantly from the actual conditions.
Existing downscaling methods primarily rely on the coupling relationship between geostationary satellites and ERA5 variables as a condition. 
However, using brightness temperature data from geostationary satellites alone fails to comprehensively capture all the changes in meteorological variables in ERA5 maps. 
To address this limitation, we can use a wider range of satellite data to make more full use of its inversion effects on various meteorological variables, thus producing more realistic results across different meteorological variables.
To further improve the accuracy of downscaling meteorological variables at any location, we propose the \textbf{M}ulti-source \textbf{O}bservation \textbf{D}own-\textbf{S}caling Model (\textbf{MODS}).
It is a conditional diffusion model that fuses data from multiple geostationary satellites GridSat, polar-orbiting satellites (AMSU-A, HIRS, and MHS), and topographic data (GEBCO), as conditions, and is pre-trained on the ERA5 reanalysis dataset.
During training, latent features from diverse conditional inputs are extracted separately and fused into ERA5 maps via a multi-source cross-attention module.
By exploiting the inversion relationships between reanalysis data and multi-source atmospheric variables—such as brightness temperature and precipitation—MODS generates atmospheric states that align more closely with real-world conditions.
During sampling, MODS enhances downscaling consistency by incorporating low-resolution ERA5 maps and station-level meteorological data as guidance. 
Experimental results demonstrate that MODS achieves higher fidelity when downscaling ERA5 maps to a 6.25 km resolution, better preserving actual meteorological details.
\end{abstract}

\section{Introduction}
Global climate data assimilation provides data on the Earth's climate system at all scales, making more accurate weather forecasting possible~\citep{lahoz2014data,carrassi2018data,li2024land}.
Among available datasets,  ERA5 is the fifth-generation global climate reanalysis dataset released by the European Centre for Medium-Range Weather Forecasts (ECMWF), which offers extensive and reliable meteorological data~\citep{hersbach2020era5,bell2021era5}.
By assimilating diverse observational sources, including satellite remote sensing, ground-based stations, and radiosonde measurements, ERA5 reconstructs historical atmospheric states using advanced numerical weather prediction models and data assimilation techniques~\citep{munoz2021era5, jiang2021evaluation, yilmaz2023accuracy}.
% It integrates a variety of observational data, such as those from satellite remote sensing, ground-based meteorological stations, and sounding balloons.
% Through advanced numerical weather prediction models and data assimilation techniques, it reconstructs the past atmospheric states~\citep{munoz2021era5, jiang2021evaluation, yilmaz2023accuracy}.

\begin{wrapfigure}{r}{0.5\columnwidth}
\vspace{-1cm}
    \centering
\includegraphics[width=0.5\columnwidth]{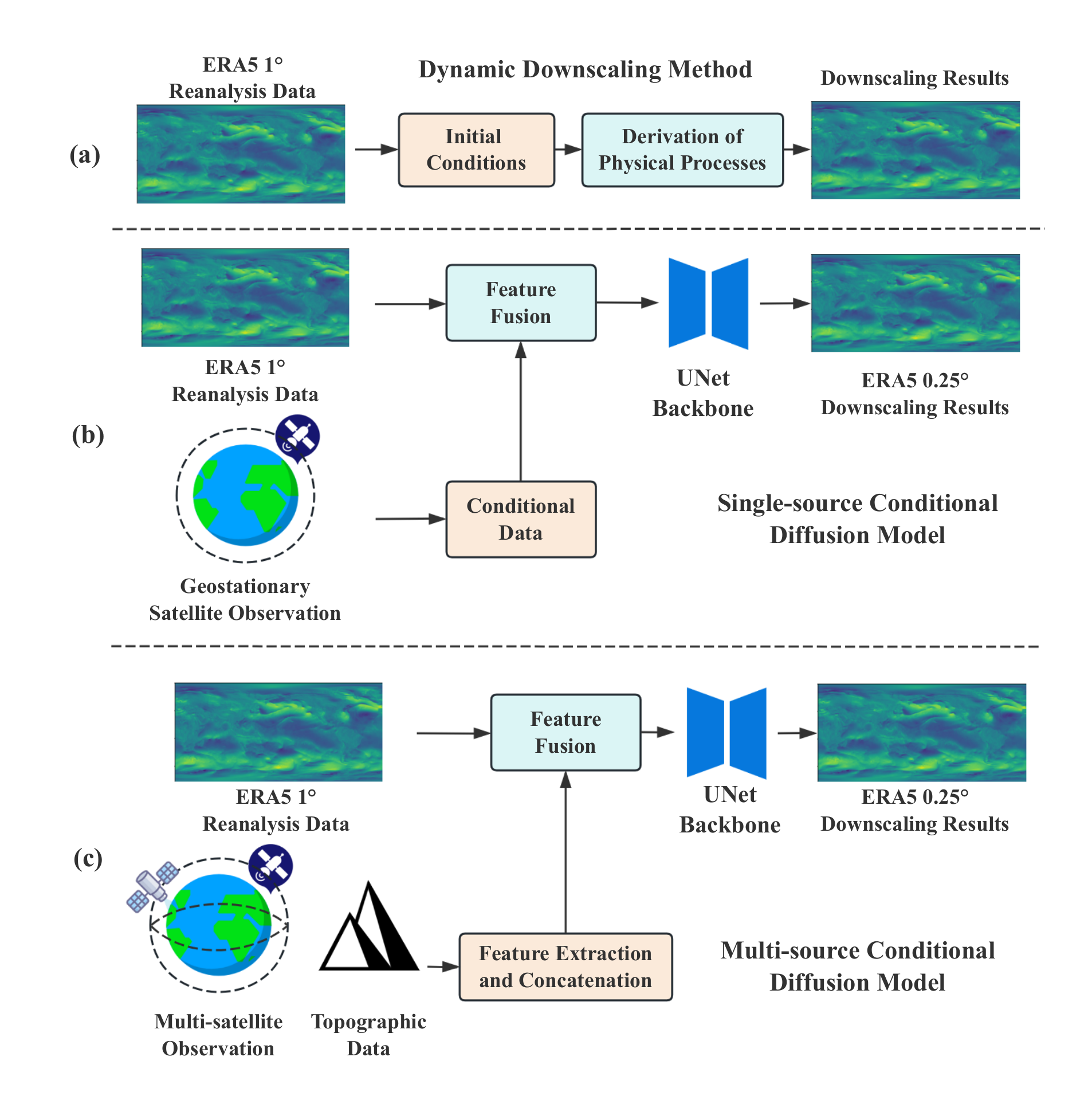}
    \vspace{-0.9cm}
    % \caption{Comparison of \textbf{(a)} Dynamic Downscaling Method, \textbf{(b)} Single-source Conditional Diffusion Model, and \textbf{(c)} Multi-source Conditional Diffusion Model. Traditional dynamic downscaling provides conditions for high-resolution regional models using global low-resolution models, capable of simulating more meso-and small-scale features. However, it requires substantial computational resources, with the computational volume increasing exponentially as the resolution rises. These multi-source data can complement each other, thus combining with the atmospheric data in ERA5 to more accurately describe the atmospheric state.}
    \caption{A paradigm shift from dynamic downscaling to multi-satellite observations conditional downscaling. Our MODS is illustrated in (c).}
    \label{fig:teaser}
\vspace{-0.3cm}
\end{wrapfigure}

% The mixed downscaling method, which involves integrating real-time observational data into the downscaling process, is a crucial approach to solving this problem, as it enhances the resolution of meteorological datasets, thereby enabling a more accurate representation of atmospheric conditions.
However, higher-resolution ERA5 data are necessary for more precise weather forecasting~\citep{cheon2024advancing, lopes2024evaluation}.
The standard ERA5 datasets, with a spatial resolution of approximately 27.75 km, lack the granularity required to accurately represent meteorological conditions in localized regions~\citep{chattopadhyay2022towards}.
Traditional physics-based downscaling algorithms entail significant computational resource requirements. 
Due to this limitation, most observational data are discarded during the assimilation process, which in turn limits the diversity of observational data integrated into the assimilation process~\citep{vandal2024global, juan2023data, giladi2021physics}.
In recent years, numerous studies have shown that artificial intelligence (AI) models have surpassed traditional physics-based downscaling methods, achieving significant improvements in both spatial detail resolution and accuracy~\citep{mardani2025residual,tomasi2024can,ling2024diffusion}.
As shown in ~\Cref{fig:teaser}, current AI-based downscaling methods frequently concentrate solely on the images themselves, relying on intricate network architectures to learn the relationships between different resolutions~\citep{behfar2024drought, koldunov2024emerging}.
However, directly generating high-resolution ERA5 maps through such methods often neglects the coupling relationship between ERA5 data and satellite observations, as the data in observation affects the atmospheric state in ERA5 maps, resulting in ERA5 maps that hardly conform to the real meteorological conditions of the atmosphere~\citep{knapp2018gridded}.

To introduce satellite observations, the pioneering work of SGD~\citep{tu2025satellite} demonstrates that a downscaling model conditioned on satellite data—while explicitly learning the coupling relationships between these observations and ERA5 maps—can produce high-resolution (HR) ERA5 representations with improved accuracy and greater fidelity to real-world meteorological conditions.
This is because ERA5 integrates a wide variety of historical observational data, particularly satellite measurements, within its advanced data assimilation and modeling systems to generate more accurate atmospheric estimates.
Therefore, the atmospheric states derived from these observational sources and data from multiple origins can also be used to evaluate the generation accuracy of generative models related to ERA5 maps~\citep{allen2025end}.
On the one hand, apart from the GridSat data, the satellite observational datasets AMSU-A, HIRS, and MHS conduct atmospheric observations from different perspectives, such as microwave and infrared. 
The provided data can complement each other, enabling a more comprehensive description of the atmospheric state.
Meanwhile, the atmospheric elements such as temperature and water vapor observed by AMSU-A, HIRS, and MHS are also crucial parameters influencing atmospheric physical processes. 
These observational data can be integrated with the atmospheric data in ERA5 to form a more accurate description of the atmospheric state~\citep{duncan2022assimilation, wylie2005trends}. 
On the other hand, existing studies~\citep{jiao2021evaluation,simmons2020global} have shown that the accuracy of ERA5 reanalysis products is closely associated with factors such as topography and climate classification. 
This also reveals that the accuracy of ERA5 reanalysis data is closely related to meteorological variables beyond topography and brightness temperature.
Therefore, more diverse and accurate satellite observational results and more precise ERA5 maps complement each other.
% The various meteorological variables observed by satellites influence the data derivation process of the ERA5 reanalysis dataset. 
% Based on this, the coupling relationship among ERA5 reanalysis data, satellite observations, and topographic data also makes it possible to use satellite observations as a condition for the condition diffusion-based downscaling method to generate HR ERA5 maps that are more in line with reality. 

% \begin{figure}[t]

Based on this, we propose MODS, a multi-source conditional diffusion-based downscaling model to construct variables such as brightness temperature and water vapor in the atmosphere from multi-angle data of different sources.
During the training process of the diffusion model, data obtained from various sources, including geostationary satellite observation, polar orbit satellite observation, and topography data, undergo feature extraction by a pre-trained encoder, and then the features are fused using the cross-attention module.
Each type of data is a different high-dimensional tensor with spatial, temporal, and channel dimensions. 
This tensor reflects a diverse range of information, including the brightness temperature of the Earth’s atmosphere, water vapor content, and surface topography.
Apart from static topographic data, these data provide a supplementary view of the atmosphere and enable the model to learn complex relationships across spaces and modalities, ensuring the authenticity of the model's generation results.
In the sampling process, we use low-resolution (LR) ERA5 maps to provide guidance for the generation process.
We utilize convolution to simulate the resolution conversion process, aiming to make the generated HR ERA5 maps match their corresponding LR ERA5 maps after the upscaling convolution simulation, thus guaranteeing the matching of details. 
Our contributions are threefold:
\begin{itemize}
\vspace{-0.3cm}
\setlength{\itemsep}{0pt}
\setlength{\parskip}{0pt}
\item In the conditional diffusion model, we have taken into account the coupling relationships existing between various satellite observational data and ERA5 maps.
By using data such as brightness temperature, water vapor, and topography as conditions, the generated HR ERA5 maps are made to better conform to real-world meteorological scenarios.
\item Effective feature extraction and fusion have been carried out on each meteorological data through pre-trained modules and cross-attention modules.
The experimental results have verified that MODS is capable of generating higher-quality ERA5 maps.
\item The sampling process of MODS can utilize diverse guidance to direct the generation process. 
Dynamic control of the generation can be achieved through different generation requirements, either via station-guided methods or reconstruction loss, which endows it with strong flexibility and practicality.
\end{itemize}

\section{Related Works}
\textbf{Satellite-assisted Data Assimilation. }
Meteorological data assimilation is an analysis technique aiming to generate a more accurate and comprehensive representation of the Earth system, which enhances the precision of numerical simulations and forecasts in various fields like meteorology and environmental science~\citep{valmassoi2023current,wang2024prediction,cluzet2022propagating}.
Nowadays, utilizing satellite observation in assimilation has been attracting increasing attention~\citep{toth2024satellite}. 
Previous study~\citep{yang2023vertical} shows that integrating satellite radiance data into regional forecasting systems or assimilation work can provide a more comprehensive understanding of atmospheric conditions. 
However, traditional assimilation methods are merely confined to meteorology-related physical approaches.
The emergence of diffusion models has pioneered new methods for satellite-assisted data assimilation~\cite{gong2024postcast,fei2023generative,tu2024taming,xu2024icediff}. 
SGD~\citep{tu2025satellite} is the first diffusion-based method to incorporate brightness temperature data from the GridSat satellite observation as condition in meteorological data assimilation tasks. 
However, considering only brightness temperature data neglects the influence of water vapor and topography data on the partial states of the atmosphere, failing to fully reflect the evolutionary characteristics of each variable in the ERA5 maps. 
Therefore, MODS combines geostationary satellites and multiple polar-orbiting satellites as conditions to assist in the generation of high-quality ERA5 meteorological variables. 

\textbf{Meteorological Downscaling. }
Downscaling aims to generate HR data from their LR counterpart, bridging resolution gaps in areas such as climate modeling and remote sensing~\citep{2024Machine,2024Erratum}. 
Traditional regression-based statistical methods leverage spatial correlations but struggle with complex nonlinear patterns~\citep{2010Downscaling}. 
Recently, deep learning methods have revolutionized this field by learning hierarchical features directly from data~\citep{2024TerraWind}.
The Generative Adversarial Networks (GANs) introduced adversarial training to generate small-scale maps that the discriminator cannot distinguish from real HR maps~\citep{2023Effects}. 
The emergence of diffusion models, with their powerful map-generation capabilities, has provided a series of potent methods for meteorological downscaling~\citep{2025Residual,2024Wind}. 
SR3~\citep{saharia2021imagesuperresolutioniterativerefinement} progressively refining LR inputs into HR outputs through a Markov chain. 
Latent Diffusion~\citep{rombach2021highresolution} reduced computational costs by operating in a compressed latent space.
However, when dealing with the ERA5 downscaling task, these methods overlook that the ERA5 data are derived from the inversion of multiple satellite observational sources, unable to generate high-quality HR ERA5 maps that are consistent with actual meteorological conditions. 
\section{Methodology}
\begin{figure}[t]
    \centering
\includegraphics[width=\linewidth]{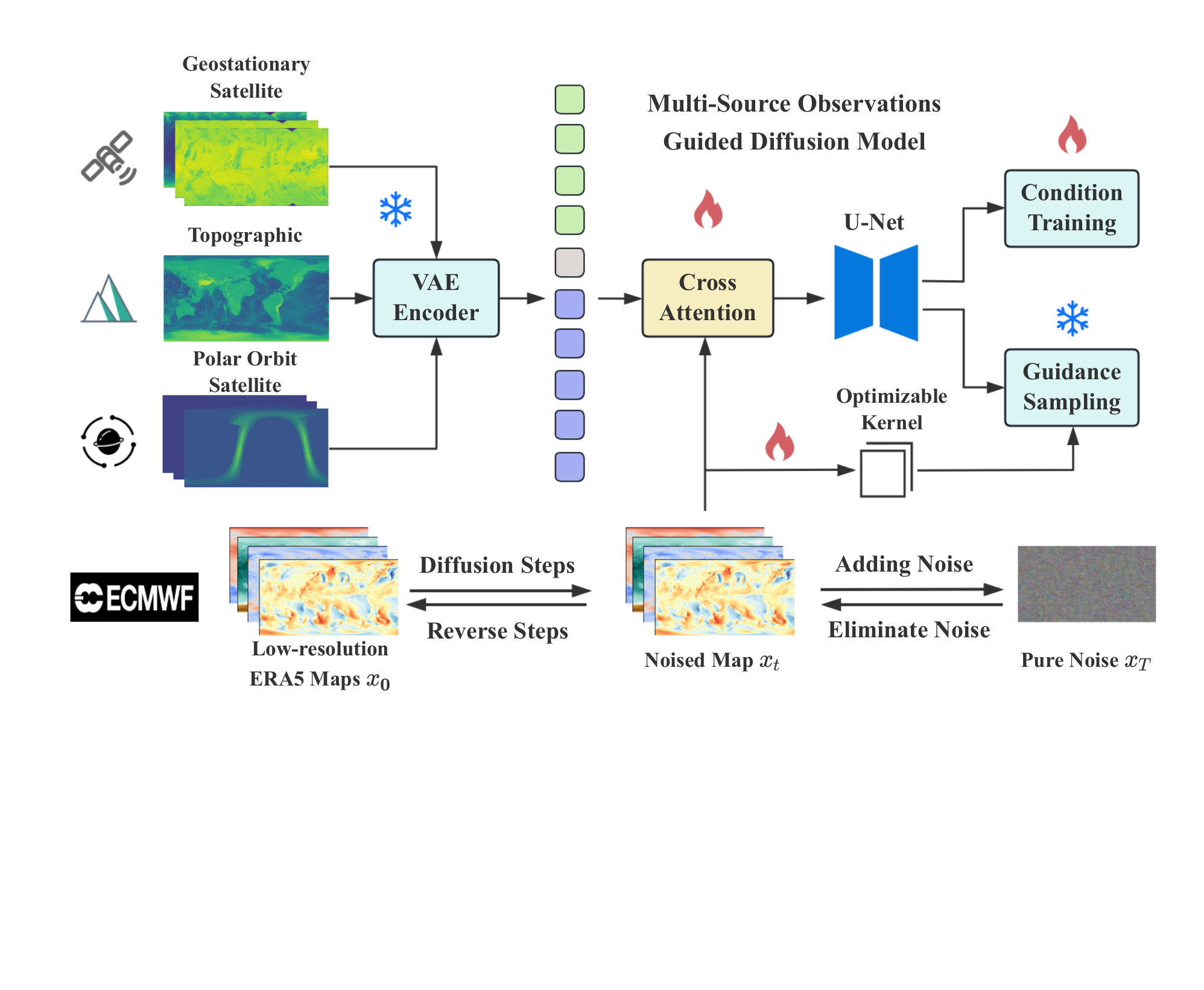}
    \vspace{-4.2cm}
    \caption{Overall Structure of MODS:
% The lower part of the image depicts the diffusion steps of ERA5 maps, and the results from each step are utilized in the conditional training and guided sampling processes in the upper part. 
During the training process, data from multiple satellite observations and topographic data are subjected to feature extraction via their respective VAE-based pre-trained encoders. 
Once these features are integrated, they serve as conditional inputs and are fused with ERA5 maps through cross-attention module, enabling the training of a conditional diffusion model.
In the sampling process, LR ERA5 maps offer guidance for the reverse process. 
The HR ERA5 outcomes generated at each reverse step are upscaled using a convolutional kernel with optimizable parameters. 
Subsequently, the gradient of the distance between the upscaled results and the LR ERA5 maps is employed to update the mean for the subsequent sampling step.
Simultaneously, station-scale data can be incorporated to assess the real-time accuracy of the generated HR ERA5 maps. 
This incorporation aids in guiding the generation process of the diffusion model from diverse perspectives, consequently improving the quality of the model's downscaling results. }
    \label{fig:flow_chart}
\vspace{-0.2cm}
\end{figure}

% \subsection{Overall Structure}
As shown in~\Cref{fig:flow_chart}, MODS aims to construct a conditional diffusion model by leveraging the coupling relationships between multi-dimensional data such as brightness temperature, water vapor, and topography data from multi-satellite observations and ERA5 maps. 
This is to ensure that the HR ERA5 maps generated through downscaling are more consistent with the actual meteorological conditions.
~\Cref{pre-trained encoder} elaborates the specific structure and function of the pre-trained encoder used to extract features from polar-orbiting satellite, geostationary satellite, and topography data.
~\Cref{conditional training} presents and analyzes the feature fusion mechanism in the conditional diffusion model.
During the sampling process,~\Cref{guided sampling} elaborately derives the implementation method of MODS, which utilizes LR ERA5 maps and station-scale meteorological variables as guidance. Meanwhile,~\Cref{guided sampling} analyzes the simulation effect of the convolutional kernel with optimizable parameters on the resolution conversion process and the methods of its parameter updating. 

\subsection{Multi-source Observation Encoder}
\label{pre-trained encoder}
\begin{figure}[h]
    \centering
\includegraphics[width=\linewidth]{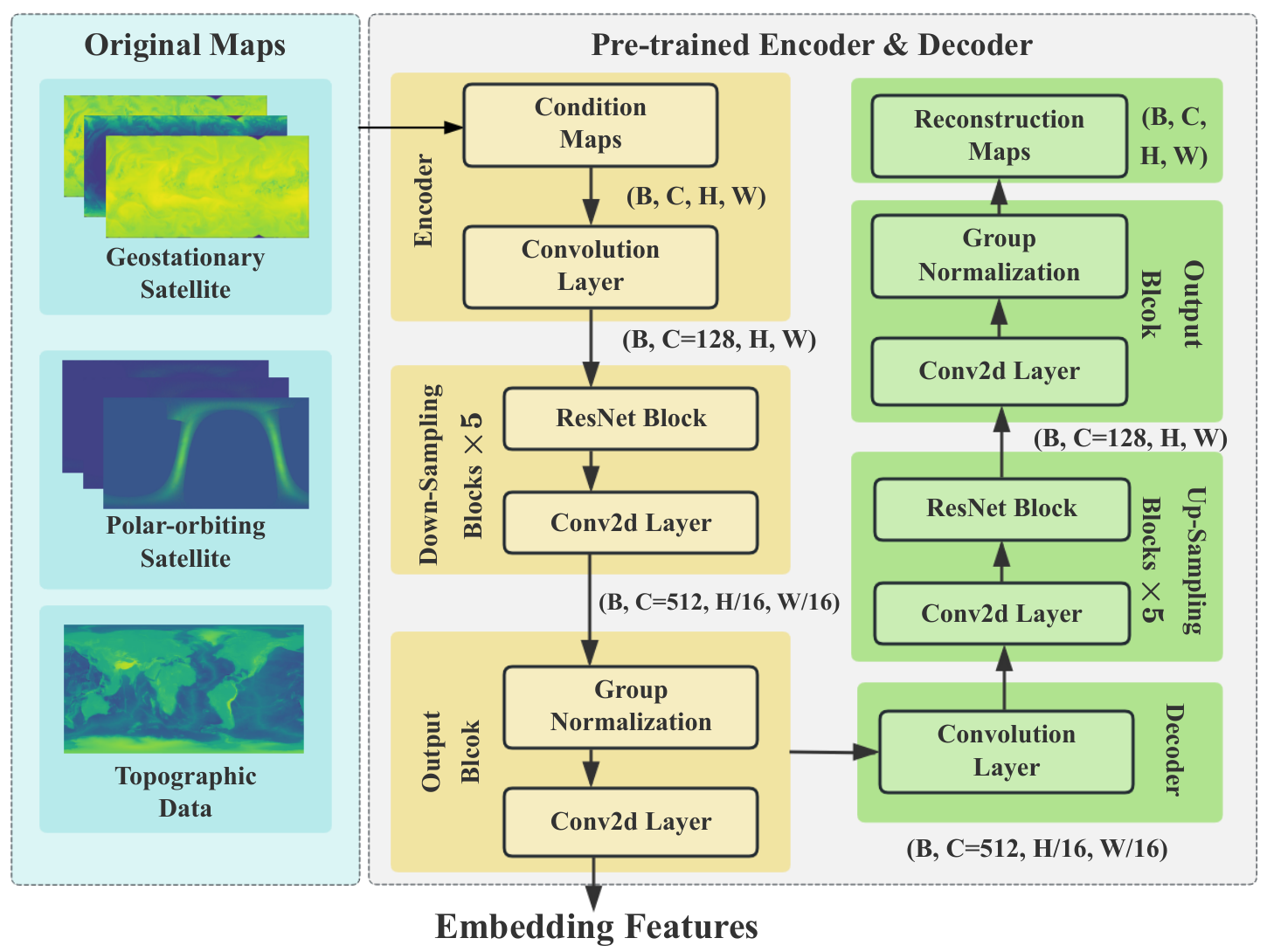}
    % \vspace{-3.2cm}
    \caption{Overall structure of the pre-trained encoder and decoder module utilized in MODS.}
    \label{fig:app_encoder}
% \vspace{-0.5cm}
\end{figure}

% Prior to MODS's application of the cross-attention module for the feature fusion of conditions, a pre-trained encoder is employed to extract the latent embedding features of diverse variables in satellite observations and topographic maps.
Before applying the cross-attention module in the MODS for conditional feature fusion, a pre-trained encoder is employed to extract the latent embedding features of various variables from satellite observations and topographic maps.
The encoder extracts the features of each channel within the condition maps via three key components: the convolution layer, the down-sampling blocks, and the output blocks, while the decoder adopts a symmetric architecture to reconstruct the condition maps.
The entire VAE-based structure undergoes training by leveraging the Mean Squared Error loss between the reconstructed maps and the original condition maps.
% Once it has passed through the pre-trained encoder $f$ with parameter $\varphi$, MODS is capable of integrating each condition map into the latent space, enabling more efficient feature extraction and thereby streamlining subsequent operations such as feature fusion.
For different sources of observational data, MODS pre-trained the corresponding encoder with parameter $\varphi_i$. 
In this way, MODS can integrate each conditional map into the latent space, thereby achieving more effective feature extraction and simplifying subsequent operations such as feature fusion.
Before the multi-source conditional data is integrated into the MODS model, it is necessary to perform corresponding pre-training on each conditional data to extract its intermediate representations.
The overall structure of the pre-trained encoder and decoder is shown in~\Cref{fig:app_encoder}.

\subsection{Multi-conditioning Mechanisms}
\label{conditional training}
Diffusion models can incorporate conditional data $y$ into the diffusion process to control the generation of $x$ by training a conditional denoising autoencoder $\epsilon_{\theta}(x_t,t,y)$.
ERA5 maps are derived from satellite observation data through methods such as assimilation. 
Moreover, the meteorological variables in these maps are also influenced by data like brightness temperature from satellite observations. 
The coupling relationship between them enables the use of satellite observation data to control the generation of high-resolution ERA5 maps.
Meanwhile, in the downsampling task of ERA5 maps, if we simply train an image-to-image downscaling model as in traditional methods without adding conditional data, the low-resolution ERA5 maps lack detailed information in small-scale regions. 
This often leads to discrepancies between the generated data in detailed areas and the actual situation.

Therefore, we attempt to use multi-modal data, including satellite observation data from geostationary satellites (GEO), polar-orbit satellites (PO), and topographic (TOPO) data as conditions to control the MODS to generate HR ERA5 maps that are more consistent with the actual situation.
A cross-attention module is incorporated to the U-Net structure to fuse the conditional information and ERA5 data \citep{rombach2022high}. 
Different conditional information is integrated through concatenation after passing through pre-trained encoders, which is effective for extracting multi-source conditional information $y = \left \{ y_i \right \} $ from each data sources $y_i$ \citep{li5173993state, martin2025generative}. 
This process allows the model to fully focus on each conditional data and perform weighted fusion of features according to the strength of the correlation.

Specifically, the cross-attention module achieves feature fusion through $W_Q(X), W_K(Y), W_V(Y)$ after the softmax operation, where $Y$ is obtained by concatenating the intermediate representations of multi-modal conditional data. 
\begin{align}
\mathbf{Y}&=Concat(\varphi_a(\mathbf{y_{GEO}}),{\varphi_b}_j(\mathbf{y_{{PO}_j}}),\varphi_c(\mathbf{y_{TOPO}})),\\
\mathbf{Atten(Q,K,V)}&=Softmax(\frac{W_Q(\mathbf{X})\cdot W_K(\mathbf{Y})}{\sqrt{d}}\cdot W_V(\mathbf{Y})),
\end{align}
where $j = 1, 2, 3$ represent the data from AMSU-A, HIRS, and MHS, respectively. 
The intermediate representation of each data $\epsilon_k(y)\in \mathbb{R}^{c_k\times d\times d_{\tau_k}}, k = a, b, c$ is processed through a mapping matrix with optimizable parameters.
$W_Q(x)$ extracts the data from ERA5 maps, while $W_K(Y)$ and $W_V(Y)$ are obtained from the multi - modal conditional data.
This structure is located at the beginning of the U-Net network. 
The parameters are trained and optimized along with the diffusion model, ultimately resulting in a multi-conditional diffusion model.

\subsection{Multi-guided Sampling}
\label{guided sampling}
The fusion of conditional data during the Training phase enables MODS to generate HR ERA5 maps that better conform to the actual meteorological conditions. 
During the sampling process, the meteorological variables from LR ERA5 maps and station observations are utilized as multi-guidance to direct the generation process of HR ERA5 maps.
The utilization of LR ERA5 maps aims to ensure that the results generated by the model do not differ in details from the corresponding LR maps. 
The incorporation of station observation data aims to measure in real-time the numerical differences of the meteorological data of the generated images at each meteorological station.

To measure the differences between the above two aspects, we introduce $\mathcal{L}_{tot} = \mathcal{L}_{res}+\mathcal{L}_{dev}$ to measure the detail difference and the numerical difference of the meteorological data at the station-scale respectively.

To avoid the problem of scale mismatch between the generation space and the input space, a scale conversion function $f_1$ is introduced to convert HR ERA5 maps into their LR counterparts.
It consists of convolutions with optimizable parameters, where the stride is used to control the multiple of the scale conversion, and its parameters are also dynamically updated as the sampling progresses.
Meanwhile, to calculate the meteorological data at the station-scale of the maps in the generation space, $f_2$ is introduced to calculate the meteorological data of the ERA5 maps at the point-scale. 
Thus, the total loss $\mathcal{L}_{total}$ during the sampling process can be expressed as: 
\begin{align}
\mathcal{L}_{total}=\lambda_1 \mathcal{L}_1(f_1(\tilde{x}_0),z_l)+\lambda_2\mathcal{L}_2(f_2(f_1(\tilde{x}_0)),f_2(z_r)).
\end{align}
Here, $\tilde{x}_0$ represents the estimated value of the immediate output at each reverse step during the sampling process. 
$z={z_l,z_r}$ represents the guidance data LR ERA5 maps $z_l$ and station-scale data $z_r$ respectively. 
$\lambda_i, i = 1, 2$ serve as weights to control the degree of incorporation of the LOSS.
$\mathcal{L}_i$ acts as a distance metric, and here the Mean Squared Error (MSE) loss is used.
It is worth noting that the settings of the LOSS and the distance metric here are quite flexible. 
Different settings can be employed according to different actual requirements, and the effect of guided sampling can also be achieved. 

Previous studies~\citep{fei2023generative} have derived the formula for the conditional distribution in the guided sampling process, where $N_1$ is a constant:
\begin{align}
\log_{}{p_\theta(x_{t}|x_{t+1},y,z)}&=\log_{}{(p_\theta(x_{t}|x_{t+1},y)p(z|x_{t})) }+N_1
\end{align}
It can be seen from this that the guided sampling is reflected in the introduction of the conditional distribution $p(z|x_{t})$.
In each reverse step, the mean and variance used are as follows:
\begin{align}
\mu = \mu_\theta&(x_t,y,t)+\Sigma_\theta\cdot\nabla_{x_t}\log{}{p(z|x_t)}\\
\Sigma & = \Sigma_\theta(x_t,y,t)
\end{align}

\begin{multicols}{2}

Here, we use a heuristic algorithm to estimate the approximate value of $\nabla_{x_t}\log{}{P(z|x_t)}$ with the gradient of the LOSS with respect to $\tilde{x}_0$:
\begin{align}
\nabla_{x_t}\log{}{p(z|x_t)}\approx \nabla_{\tilde{x}_0}LOSS(\tilde{x}_0,z)
\end{align}

Based on this, the overall process of guided sampling is obtained.
As shown in \Cref{alg.1}, in each reverse step of the conditional diffusion model, the immediate output estimated value $\tilde{x_0}$ is first obtained according to the HR ERA5 maps $x_t$ in the generation space.
Subsequently, substitute it into the function $f=(f_1,f_2)$ at time $t$ to calculate the deviation $LOSS_t$ at this time.
The gradient of this deviation is utilized to update the mean of the sampling process, and sample $x_{t - 1}$ by combining the mean and variance of the noise prediction network $\epsilon_\theta$.
Meanwhile, the gradient of $Loss_{{res}_t}$, which represents the difference between the LR in the input space and the scale-converted HR ERA5 maps, is also used to update the parameters in $f_1$ so as to more effectively simulate the scale conversion in the subsequent reverse steps. 

\begin{algorithm}[H]\small
\renewcommand{\algorithmicrequire}{\textbf{Input:}}
\renewcommand{\algorithmicensure}{\textbf{Output:}}
\caption{\textbf{Multi-guided Sampling}}
\label{alg.1}
\begin{algorithmic}[1]
\REQUIRE Conditional diffusion model  $\epsilon_\theta(x_t,y,t)$ with conditional data $y=\left \{ y_i \right \},i=1,2,\dots,n $ and its pre-trained encoder $\varphi_i$. Guidance data $z=(z_l,z_s)$. Scale conversion function $f_1$ with parameter $\vartheta$. Interpolation function $f_2$. Loss weight $\lambda=(\lambda_1,\lambda_2)$. Learning rate $l$.
\ENSURE Output HR ERA5 map $x_0$. 

\STATE Sample $x_T$ from $\mathcal{N}(0,I)$

\STATE $y=Concat(\varphi_1(y_1),\varphi_2(y_2),\dots,\varphi_n(y_n))$

    \FORALL{t from T to 1}
        \STATE $\tilde{x} _0=\frac{x_t}{\sqrt{\bar{\alpha}_t }}-\frac{\sqrt{1-\bar{\alpha}_t}\epsilon_\theta(x_t,y,t)}{\sqrt{\bar{\alpha}_t }}$\
        
        \STATE $\mathcal{L}_{{res}_t} =\mathcal{L}_1(f_1(\tilde{x}_0),z_l)$\

        \STATE $\mathcal{L}_{{dev}_t} =\mathcal{L}_2(f_2(f_1(\tilde{x}_0)),f_2(z_l))$\

        \STATE $\mathcal{L}_{tot_t}=\lambda_1(\mathcal{L}_{res_t})+\lambda_2(\mathcal{L}_{dev_t})$\

        \STATE $\tilde{x}_0 \gets \tilde{x}_0-\frac{s(1-\bar{\alpha}_t) }{\sqrt{\bar{\alpha}_{t-1}}\beta_t}\nabla_{{\tilde{x}}_0}\mathcal{L}_{tot_t}$\
        
        \STATE $\tilde{\mu}_t=\frac{\sqrt{\bar{\alpha}_{t-1}}\beta_t}{1-\bar{\alpha}_t}\tilde{x}_0+\frac{\sqrt{\bar{\alpha}_{t}}(1-\bar{\alpha}_{t-1})}{1-\bar{\alpha}_t}{x}_t$\

        \STATE $\tilde{\beta}_t=\frac{1-\bar{\alpha}_{t-1}}{1-\bar{\alpha}_t}\Sigma_\theta(x_t,y,t)$\

        \STATE Sample $x_{t-1}$ from $\mathcal{N}(\tilde{\mu}_t,\tilde{\beta}_tI)$\

        \STATE $\vartheta_{t-1}=\vartheta_t-l\nabla_\vartheta(\mathcal{L}_{res_t})$\
    \ENDFOR
\STATE \textbf{return}  $x_0$
    \end{algorithmic}
\end{algorithm}

\end{multicols}

\section{Experiment}
\subsection{Data Introduction. }
\textbf{ERA5}~\citep{hersbach2020era5} is the fifth-generation global climate reanalysis dataset released by the European Centre for Medium-Range Weather Forecasts (ECMWF). 
It combines model data with satellite observations and other sources of information, and generates a comprehensive set of global atmosphere, oceans, and land data with a resolution of 0.25° through techniques like data assimilation. 
% The original spatial resolution of the data reaches 0.25°.
In the experiment, the variables we used include $U_{10}$, $V_{10}$, $T_{2m}$, and $MSL$. 
The first two variables represent the zonal and the meridional wind speed component at a height of 10 meters, respectively. 
$T_{2m}$ refers to the air temperature at a height of 2 meters, and $MSL$ stands for mean sea-level pressure.
% All of these variables are influenced by factors such as atmospheric brightness temperature. 
% More precise data at a smaller scale can contribute to more accurate decision-making in human activities and weather forecasting. 
\textbf{Conditional Data. }
\Cref{app:dataset} presents information such as the types and sizes of data integrated as conditional data into the condition diffusion model.

\textbf{GridSat} dataset~\citep{skofronick2015global} is derived from the observational data in visible, infrared and other bands collected by multiple meteorological satellites from the NOAA series. 
These data are then comprehensively processed and gridded. 
The data content records meteorological parameters such as cloud-top temperature and radiance in a regular grid format. 
The cloud-top temperature reflects the thermal state, while the radiance reflects the radiation characteristics of the Earth's surface and the atmosphere.

\textbf{HIRS} (High-Resolution Infrared Radiation Sounder)~\citep{shi2011three} is an instrument installed on meteorological satellites, which is used to measure the radiation of the Earth's atmosphere in the infrared band. 
The data obtained by these instruments are processed to form the HIRS dataset. 
The instrument data mainly provides information such as the vertical temperature profile of the atmosphere. 
By measuring infrared radiation of different wavelengths and using the principle of radiative transfer, the temperature and water vapor conditions at different altitudes in the atmosphere can be retrieved.

\textbf{MHS} (Microwave Humidity Sounder)~\citep{bonsignori2007microwave} is installed on meteorological satellites and is used to measure the water vapor distribution in the atmosphere. 
The MHS instrument utilizes the absorption characteristics of water vapor in the microwave band and measures microwave radiation to retrieve the water vapor content in the atmosphere.

\textbf{AMSU-A} (Advanced Microwave Sounding Unit-A)~\citep{mo1996prelaunch} is a microwave detection instrument installed on meteorological satellites, which is mainly used to measure the temperature distribution of the atmosphere. 
By measuring microwave radiation at different frequencies, it can obtain temperature information at different altitudes in the atmosphere.

\textbf{GEBCO} (General Bathymetric Chart of the Oceans)~\citep{mayer2018nippon} terrain dataset is sourced from measurements by ship-borne echo sounders, satellite altimetry data, etc. 
The data includes global DEM data from grid-scale to watershed-scale. 
It combines grid DEM with high-resolution satellite remote-sensing images and divides the global land and ocean areas. 

\begin{table}[t]
\centering
\caption{Data introduction of various conditional datasets.}
% \vspace{-0.3cm}
\resizebox{\linewidth}{!}{
\begin{tabular}{c|c| c | c c}
    \toprule[1pt]
    Dataset & Type & Acquisition & Original Shape & Intermediate Representation Shape \\
    \midrule
    HIRS & \multirow{4}{*}{Meteorological Satellite} & Infrared Remote Sensing & (20,960,1920) & (160,60,120) \\
    MHS &  & Microwave Detector & (5,960,1920) & (40,60,120) \\
    AMSU - A &  & Microwave Detector & (15,960,1920) & (120,60,120) \\
    GridSat &  & Satellite Observation & (3,960,2560) & (24,60,160) \\
    \midrule
    GEBCO & Topography & Echo Probing \& Satellite Altimetry & (1, 1, 1296, 2592) & (16, 60, 160) \\
    \bottomrule[1pt]
\end{tabular}
}
\label{app:dataset}
\vspace{-0.2cm}
\end{table}

\textbf{Implementation Details.}
For the pre-trained encoder, we pre-train it using AMSU-A, MHS, GridSat, and HIRS data with a 6-hour interval from 2010 to 2023, with a time error within 30 minutes.
During pre-training, the Mean Squared Error between the reconstruction results and the original observed maps is utilized as the training loss, and the number of epochs used is 10.
Regarding the topographic data GEBCO, since the differences in topographic data across different time periods are not significant, we use single-frame data for pre-training.
For the training process of MODS, the hourly ERA5 maps from 2010 to 2021 with a shape of (720×1440) are used as the training set, the data from 2021 to 2022 as the validation set, and the data from 2023 as the test set.
All the training tasks are conducted on an NVIDIA A100 80GB GPU.
The entire training process consists of 200,000 steps and lasts for 2-3 days.
MODS is optimized using AdamW with $\beta_1 = 0.9$ and $\beta_2 = 0.999$ in 16-bit precision with loss scaling, while maintaining 32-bit weights, Exponential Moving Average (EMA), and optimizer state. 
Each layer of the U-Net contains 2 residual blocks, and the UNet structure has a total of 4 layers.
Meanwhile, the diffusion variance $\beta_t$ is set as hyperparameter increasing linearly from $\beta_1 = 10^{-4}$ to $\beta_T = 0.02$.
During the sampling process, the kernel of the scaling transformation function is $9\times 9$, which is implemented utilizing group convolution. 

\textbf{Evaluation Metrics. }
We utilized station-scale data to examine the differences between the values of all four variables in the downscaled ERA5 maps and the actual meteorological values at various stations on a global scale.
The actual meteorological variable data at the stations were sourced from the Weather 5k dataset~\citep{han2024weather5k}. 
This dataset is a large-scale time-series dataset for sparse weather forecasting, which encompasses 5,672 weather stations distributed across the globe.
The comparison metrics included the Mean Absolute Error (MAE) and the Mean Squared Error (MSE), which were employed to quantify the discrepancies between the output results and the actual values.
The methods compared with MODS included off-the-shelf interpolation-based methods and diffusion-based methods.
We extracted the values of variables at the weather 5k stations from the small-scale ERA5 maps of different methods by using the $grid\_sample$ function in PyTorch. 

\subsection{Main Results}

\begin{table}[t]\small
  \centering
  \caption{Results of MODS and off-the-shelf downscaling methods on $U_{10}$, $V_{10}$, $T_{2M}$ and $MSL$. Different colors represent different types of methods. A mixture of the LR ERA5 maps and station-scale data is utilized as multi-guidance in MODS.}
  % \vspace{-0.3cm}
  \label{tab:main}
  \resizebox{0.9\linewidth}{!}{
  \begin{tabular}{c|c c |c c |c c |c c }
    \toprule[1pt]
     \multirow{2}{*}{Methods}&\multicolumn{2}{c|}{$U_{10}$}&\multicolumn{2}{c|}{$V_{10}$}&\multicolumn{2}{c|}{$T_{2m}$}&\multicolumn{2}{c}{$MSL$}\\
     \cmidrule(lr){2-9}
     
    &MSE&MAE&MSE&MAE&MSE&MAE&MSE&MAE\\
    \midrule
    \rowcolor{green!10}
ERA5 $1^{\circ} $  & 53.18 & 5.95 & 38.51 & \textbf{4.95} & 216.27 & 11.39 & 470.06 & 15.78 \\
    \rowcolor{green!10}
Bilinear  & 56.55 & 6.16 & 37.94 & 5.03 & 198.40 & 11.26 & 464.51 & 16.05 \\
    \rowcolor{green!10}
Bicubic  & 55.74 & 6.11 & \textbf{37.88} & 4.98 & 201.03 & 11.15 & 458.82 & 15.97 \\
    \rowcolor{gray!10}
SwinRDM ~\citep{chen2023swinrdm} & 53.18 & 5.95 & 38.51 & \textbf{4.95} & 216.27 & 11.39 & 470.06 & 15.78 \\
    \rowcolor{gray!10}
Ref-SR ~\citep{huang2022task}& 62.72 & 6.15 & 43.12 & 5.17 & 195.42 & 11.02 & 395.42 & 15.10 \\
    \rowcolor{gray!10}
$C^2$-Matching ~\citep{jiang2021robust}& 65.12 & 6.02 & 44.57 & 5.41 & 200.17 & 11.36 & 410.72 & 15.32 \\
    \rowcolor{blue!10}
DGP ~\citep{pan2021exploiting} & 97.02 & 7.96 & 47.52 & 5.07 & 214.58 & 11.94 & 529.72 & 18.54 \\
    \rowcolor{blue!10}
GDP ~\citep{fei2023generative} & 94.99 & 7.85 & 40.17 & 5.04 & 190.11 & 10.82 & 511.71 & 18.14 \\
    \rowcolor{blue!10}
DDNM ~\citep{wang2022zero} & 52.08 & 5.94 & 42.17 & 5.65 & 193.24 & 11.77 & 396.58 & 15.12 \\
    \rowcolor{blue!10}
HyperDS ~\citep{liu2024observation}& 53.72 & 6.01 & 41.37 & 5.26 & 191.83 & 10.87 & 384.72 & 14.72 \\
    \rowcolor{blue!10}
SGD ~\citep{tu2025satellite} & 52.03 & 5.93 & 39.82 & 5.05 & 187.69 & 10.63 & 374.39 & 14.49 \\
\midrule
% MODS A& 60.84 & 7.01 & 45.86 & 6.40 & 201.87 & 11.21 & 408.85 & 15.32 \\
    \rowcolor{blue!10}
MODS& \textbf{43.43} & \textbf{5.36} & 39.96 & 5.05 & \textbf{155.64} & \textbf{9.42} & \textbf{371.28} & \textbf{14.20} \\
% MODS C& 57.52 & 6.08 & 41.02 & 5.84 & 180.85 & 10.47 & 373.24 & 14.68 \\
    \bottomrule[1pt]

  \end{tabular}
  }
\vspace{-0.2cm}
\end{table}

\begin{figure}[t]
    \centering
\includegraphics[width=\linewidth]{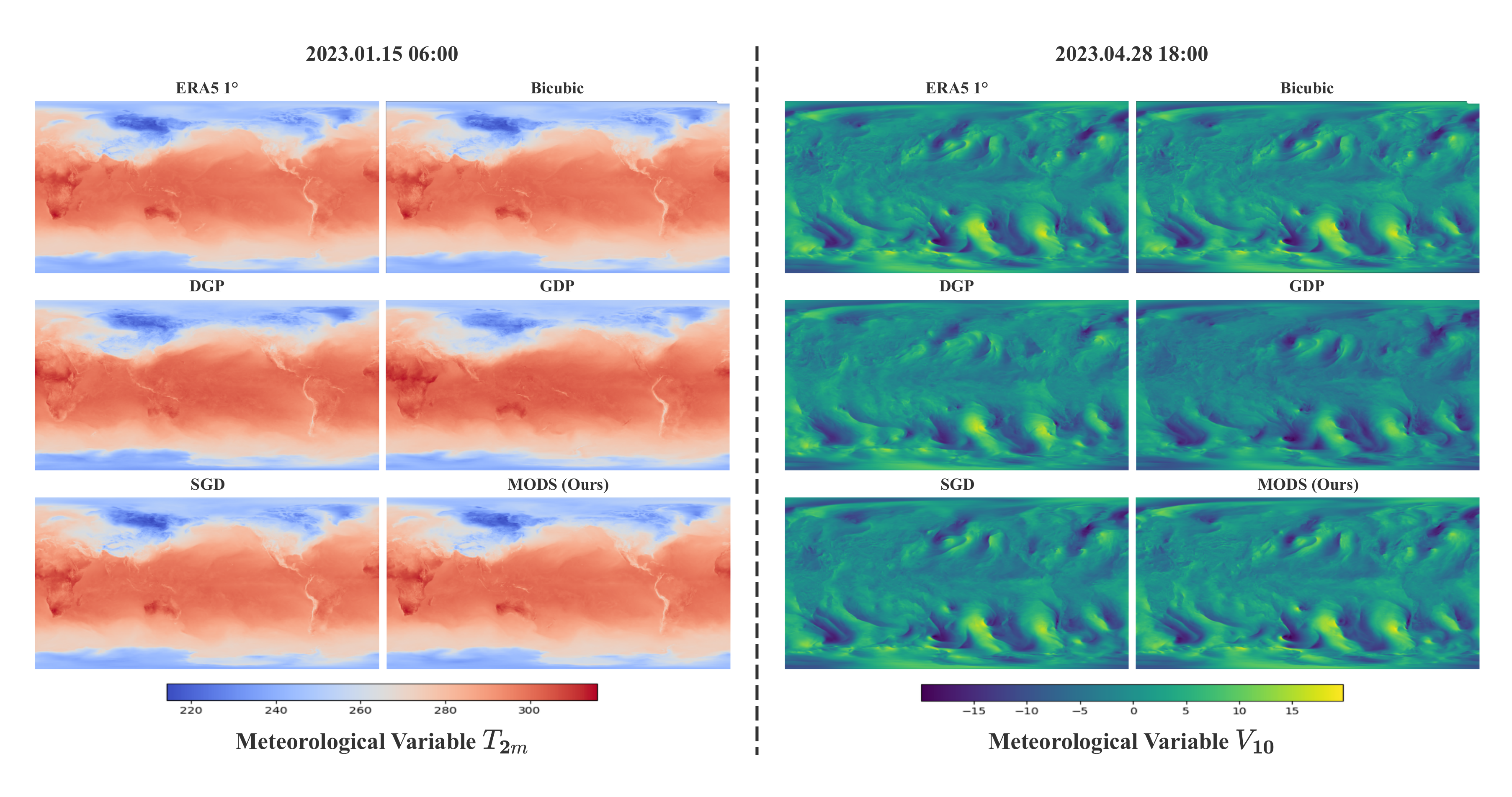}
    \vspace{-0.9cm}
    \caption{Comparison of the downscaling results of ERA5 maps obtained by MODS with other methods such as interpolation-based and diffusion-based methods.}
    \label{fig:compare}
% \vspace{-0.2cm}
\end{figure}

\begin{figure}[t]
    \centering
    \vspace{-0.5cm}
\includegraphics[width=\linewidth]{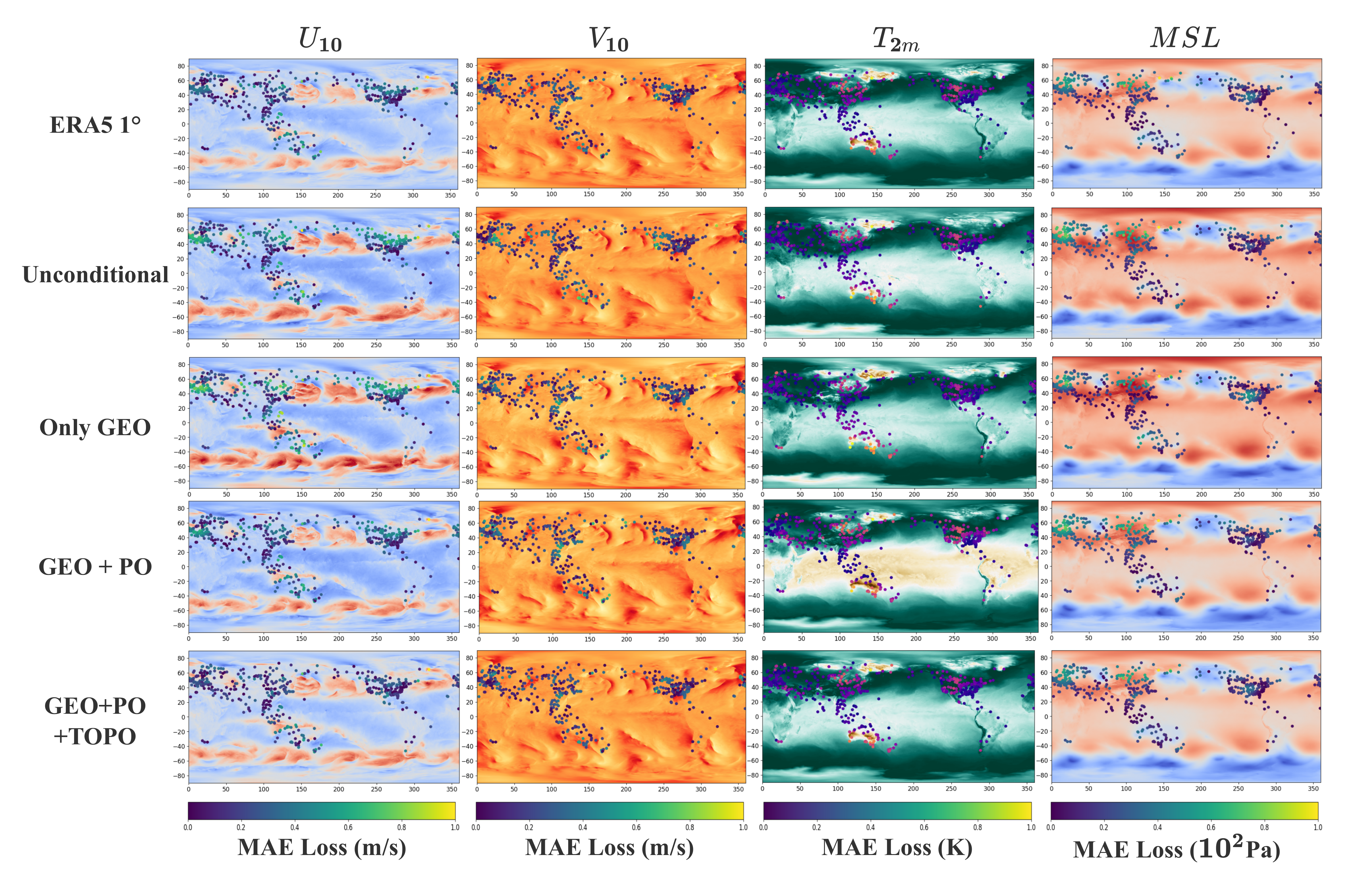}
    \vspace{-0.9cm}
    \caption{Comparison of meteorological differences at the station-scale when using different data as conditions. The color of each point represents the value difference between the results and the real meteorological data at the Weather 5K stations. Brighter colors indicate larger value differences.}
    \label{fig:ablation}
\vspace{-0.2cm}
\end{figure}

To measure the difference between the results of the downscaling task of MODS on ERA5 maps and real-world meteorological scenarios, we compare the metrics and results with off-the-shelf methods.
Existing methods include interpolation-based methods, diffusion-based methods, and other methods that can be utilized to solve super-resolution or down-scaling tasks.
It is difficult to quantify whether high-resolution ERA5 maps conform to actual meteorological scenarios through intuitive image comparison. 
Therefore, we utilize the Weather 5K dataset to measure the differences in meteorological variables at the station scale.
As shown in ~\Cref{tab:main}, MODS outperforms existing methods in all metrics of $U_{10}$, $T_{2M}$, and $MSL$, and also shows certain improvements compared with direct interpolation methods.
Notably, compared with SGD, which only uses single-source GridSat satellite observation data as a condition, MODS has advantages in all metrics.
In particular, the improvement in the $T_{2M}$ metric, which reflects temperature data, is particularly significant. 
This verifies that when multi-source observation data is used as a condition, brightness temperature data from different sources can effectively enhance the model's ability to accurately generate the meteorological variable of temperature.

\Cref{fig:compare} visually presents a comparison of images for various variables in the downscaled ERA5 maps at multiple time stamps.
Compared with off-the-shelf methods, MODS exhibits a lower degree of detail loss. 
This is attributed to the use of LR ERA5 maps as guidance to dynamically control the details in each step of the reverse process.

\subsection{Comparison of Reconstruction Results}
\begin{figure}[h]
    \centering
\includegraphics[width=\linewidth]{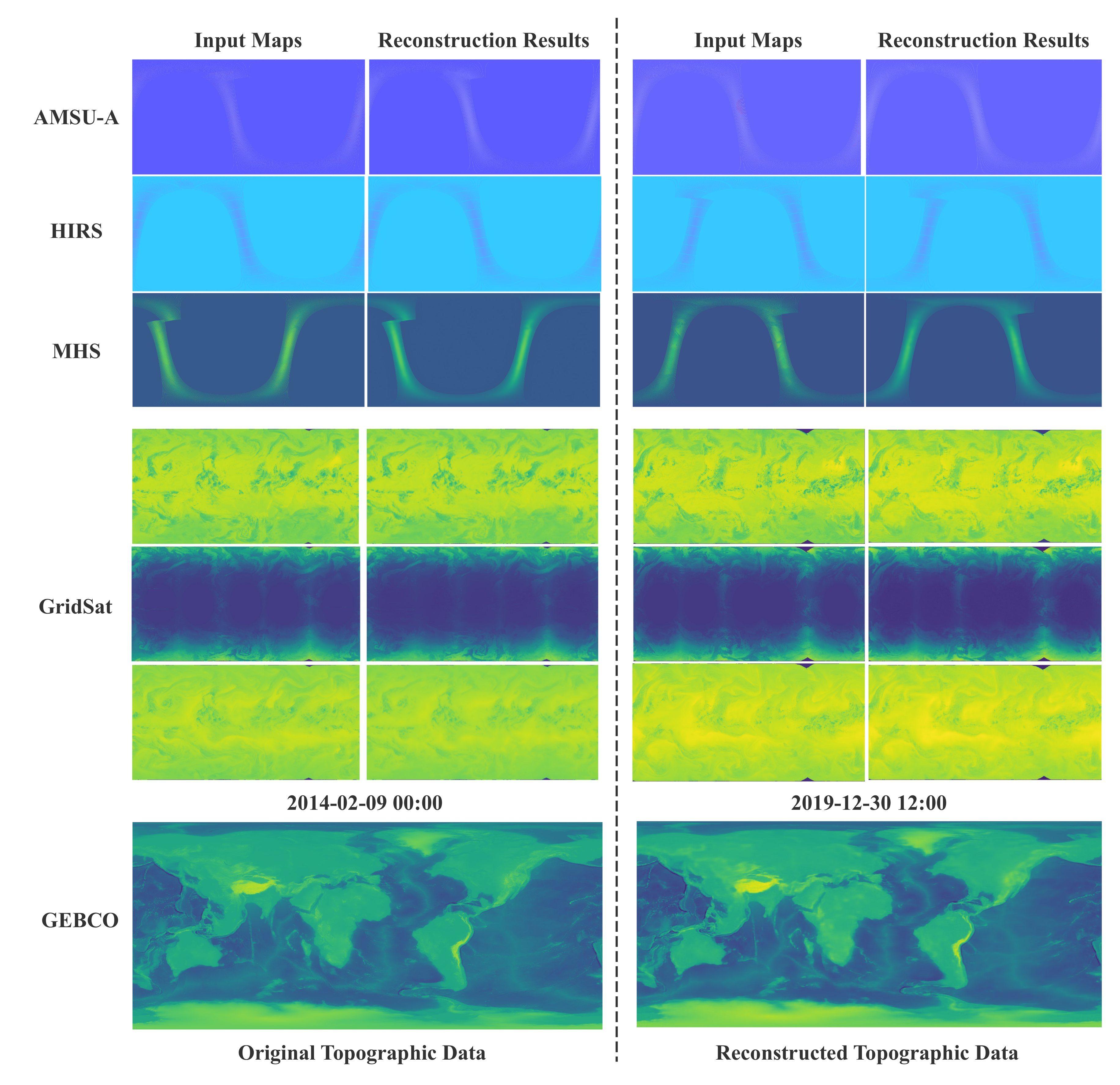}
    % \vspace{-3.2cm}
    \caption{Comparison of input and MODS reconstruction maps for different data channels.}
    \label{fig:app_res}
% \vspace{-0.5cm}
\end{figure}

% Before the cross-attention module is utilized for feature fusion, it is necessary to perform corresponding pre-training on each conditional data to extract its intermediate representations.
To verify the effectiveness of the pre-trained model, we compared the reconstruction results of the model with the original observed maps.
As shown in~\Cref{fig:app_res}, there are no discernible differences in details between the reconstruction results of the pre-trained models for each satellite observation and topographic data and their respective inputs.
Meanwhile, there is no significant degradation in quality, such as blurring. 
This indicates that the pre-training is relatively sufficient, and the existing model can achieve effective feature extraction and reconstruction.

\subsection{Comparison Regarding the Influence of Different Conditional Data}
\begin{figure}[t]
    \centering
\includegraphics[width=\linewidth]{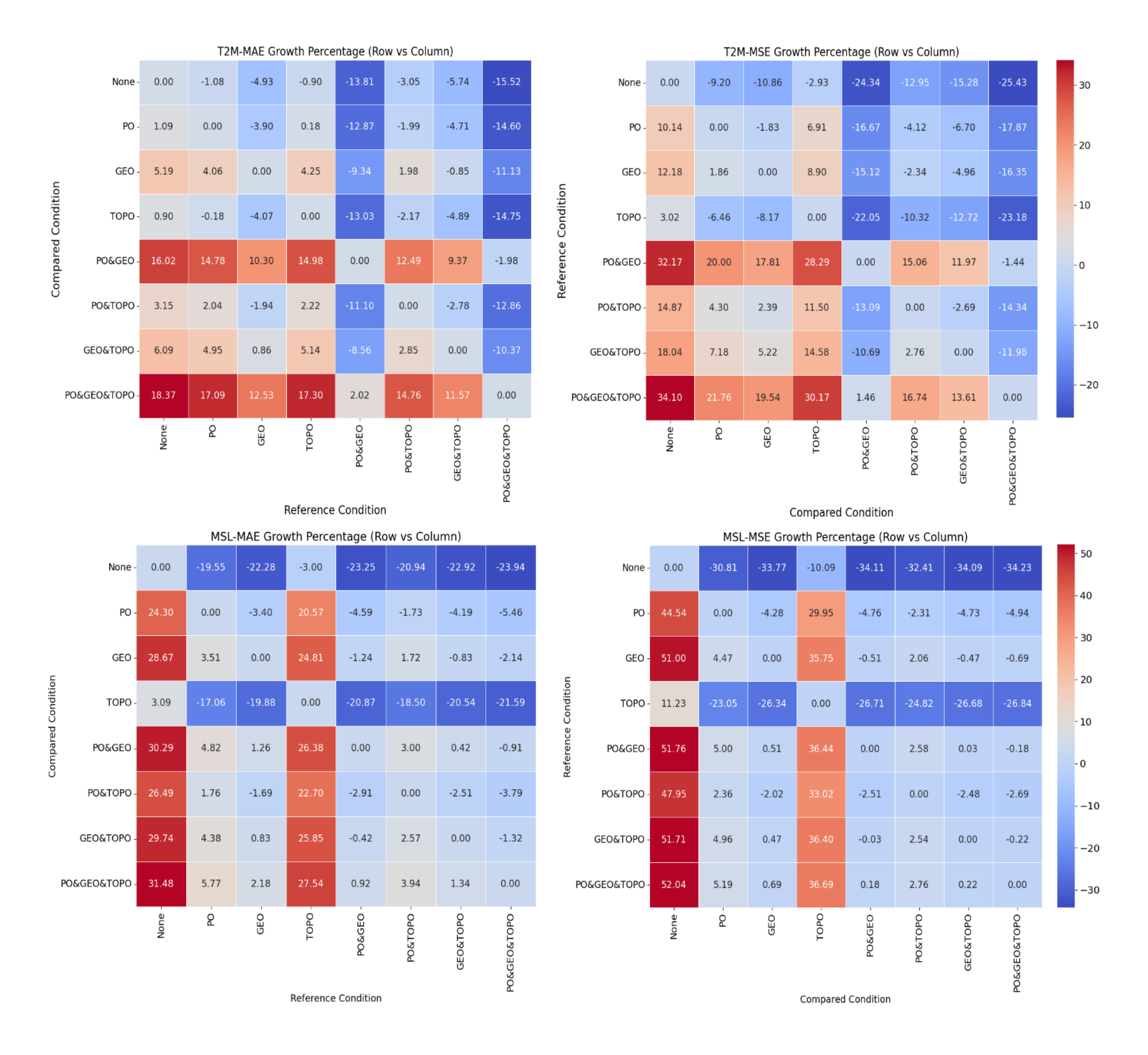}
    % \vspace{-3.2cm}
    \caption{Heatmap figure of the relative error losses of different conditional data for variables $T_{2M}$ and $MSL$.
The horizontal and vertical axes represent the types and quantities of introduced conditions.
The values in the heatmap represent the percentage improvement of the MSE and MAE results of the model under the horizontal-axis conditions for this variable compared with those under the vertical-axis conditions. }
    \label{fig:app_condition}
% \vspace{-0.5cm}
\end{figure}

To validate the impact of different data in multi-conditional data on the downscaling results of ERA5 maps, the main text conducts a quantitative comparison. 
% metrics comparison.
The results show that the introduction of multi-conditional data has a relatively large impact on the variable $T_{2M}$, followed by $MSL$.
The heatmap in \Cref{fig:app_condition} illustrates the influence of topographic data (TOPO), Geostationary Satellite Observation data (GEO), and Polar-orbiting satellite data (PO) on the results of these two variables.
Through comparison, it can be found that GEO data has the most significant improvement effect on the results compared with other data.
This is because the brightness temperature data in GridSat is a crucial factor affecting the atmospheric state.
Moreover, PO also has a relatively obvious impact on the results, suggesting that variables such as water vapor content influence both temperature variable $T_{2M}$ and the pressure variable $MSL$.
Interestingly, combining GEO and PO yields results comparable to those of MODS, highlighting their synergistic effect.
In contrast, topographic data (represented by GEBCO) has a comparatively minor influence, though it still contributes measurably to the downscaling results.

\subsection{Ablation Study}
\textbf{The effectiveness of MODS in different condition datasets.}
To evaluate the role of the multi-source data used by MODS, including the topographic data (TOPO), Geostationary Satellite Observation data (GEO), and Polar-orbiting satellite data (PO), we conduct an ablation study to examine the metric results when only part of this data is utilized.
Water vapor and brightness temperature data from satellite observations are factors influencing the atmospheric state. 
Multi-source data can simulate and reconstruct the real-world atmospheric state from multiple dimensions.
The results in ~\Cref{tab:ablation} validate the effectiveness of this setting. 
Compared to using only part of the multi-source data as conditions, MODS achieves ERA5 downscaling results that are closer to real-world conditions for all meteorological variables.
As shown in ~\Cref{fig:ablation}, the comparison also reveals that the brightness temperature data from GEO has a relatively greater impact on the results, especially on the corresponding $T_{2M}$ metric.
In contrast, the TOPO data has a relatively smaller but more comprehensive impact on the results.

\textbf{The effectiveness of MODS utilizing different guidance during sampling.}
During the sampling process of MODS, the weights of different parts in the loss function can be flexibly adjusted to modify the types of data added as guidance.
By incorporating LR ERA5 maps as part of the guidance, MODS can prevent the generation of HR ERA5 corresponding maps that suffer the loss of details compared to the LR ERA5 maps.
Utilizing station-scale meteorological station data enables real-time measurement and control of the differences between the generated meteorological data at various stations and the actual conditions.
By assigning different weights to the above two data sources, multi-guided sampling can be achieved.
As shown in \Cref{tab:multi_guide}, using station observations and LR ERA5 maps as multi-guidance helps MODS generate HR ERA5 maps that are more consistent with actual meteorological scenarios.
This effect is more prominent when station observations are solely used as guidance. However, an excessive focus on the meteorological data at stations may also lead to the loss of details in the generated results.

\begin{table}[t]\small
  \centering
  \caption{Ablation study on the downscaling results of MODS when using different satellite data as conditions. }
  % \vspace{-0.3cm}
  \label{tab:ablation}
  \resizebox{0.9\linewidth}{!}{
  \begin{tabular}{c|c|c c|c c|c c|c c}
    \toprule[1pt]
     \multirow{2}{*}{Methods} &\multirow{2}{*}{Conditions}&\multicolumn{2}{c|}{$U_{10}$}&\multicolumn{2}{c|}{$V_{10}$}&\multicolumn{2}{c|}{$T_{2M}$}&\multicolumn{2}{c}{$MSL$}\\
     \cmidrule(lr){3-10}

    &&{MSE}&{MAE}&{MSE}&{MAE}&{MSE}&{MAE}&{MSE}&{MAE}\\
    \midrule
    \multirow{7}{*}{{\begin{tabular}{@{}c@{}}Partial-conditional or\\unconditional MODS\end{tabular}}}
    &Uncondition&74.02&6.87&47.11&5.52&208.72&11.15&564.50&18.67\\
    &GEO&51.42&5.76&40.18&5.12&186.05&10.60&373.85&14.51\\
    &PO&53.04&5.81&40.01&5.10&189.51&11.03&390.56&15.02\\
    &TOPO&70.58&6.44&46.36&5.17&202.60&11.05&507.52&18.11\\
    &GEO \& PO&46.72&5.61&39.98&\textbf{5.03}&157.92&9.61&371.96&14.33\\
    &GEO \& TOPO&49.83&5.81&40.06&5.07&176.82&10.51&372.09&14.39\\
    &PO \& TOPO&51.94&5.78&39.96&5.05&181.70&10.81&381.54&14.76\\
    \midrule
    Multi-conditional MODS&GEO \& PO \& TOPO& \textbf{43.43} & \textbf{5.36} & \textbf{39.96} & 5.05 & \textbf{155.64} & \textbf{9.42} & \textbf{371.28} & \textbf{14.20}\\
    \bottomrule[1pt]

  \end{tabular}
  }
  \vspace{-0.2cm}
\end{table}

\begin{table}[t]\small
  \centering
  \caption{Ablation study on the different types of data as guidance in MODS sampling process. }
  % \vspace{-0.3cm}
  \label{tab:multi_guide}
  \resizebox{0.9\linewidth}{!}{
  \begin{tabular}{c|c c|c c|c c|c c}
    \toprule[1pt]
     \multirow{2}{*}{Methods}&\multicolumn{2}{|c|}{$U_{10}$}&\multicolumn{2}{c|}{$V_{10}$}&\multicolumn{2}{c|}{$T_{2M}$}&\multicolumn{2}{c}{$MSL$}\\
     \cmidrule(lr){2-9}

    &{MSE}&{MAE}&{MSE}&{MAE}&{MSE}&{MAE}&{MSE}&{MAE}\\
    \midrule

    MODS guided by LR ERA5& 60.84 & 7.01 & 45.86 & 6.40 & 201.87 & 11.21 & 408.85 & 15.32\\
    MODS guided by station-scale data& \textbf{43.43} & \textbf{5.36} & \textbf{39.96} & \textbf{5.05} & \textbf{155.64} & \textbf{9.42} & \textbf{371.28} & \textbf{14.20}\\
    Multi-guided MODS& 54.32 & 5.82 & 40.52 & 5.66 & 172.04 & 10.12 & 374.26 & 14.49\\

    \bottomrule[1pt]

  \end{tabular}
  }
  \vspace{-0.2cm}
\end{table}

\section{Limitation and Future Work}
The conditional data utilized in the MODS includes topographic data, geostationary satellite observation data, and polar-orbiting satellite observation data. However, direct observations of wind speed, an important variable, are absent from these satellite observations. For instance, incorporating satellite data from ASCAT, which is related to wind speed, may further enhance the performance of MODS in estimating wind-speed-related variables such as $U_{10}$ and $V_{10}$. 

\section{Conclusion}
The proposed model MODS is a downscaling model for ERA5 data based on a multi-source conditional diffusion model.
It integrates three different types of polar-orbiting satellite data, including AMSU-A, HIRS, and MHS, along with geostationary satellite data GridSat and topographic data GEBCO as multi-source conditions.
The brightness temperature, water vapor, and topographic data in the multi-source data are key parameters influencing the physical state of the atmosphere. 
Moreover, the data from different sources can complement each other, enabling a more comprehensive description of the atmospheric state.
Meanwhile, the ERA5 data also incorporates satellite observation data in its assimilation system to achieve a more accurate estimation of the atmospheric state.
Based on this, during the training process, the multi-source data undergoes feature extraction through a pre-trained encoder structure. 
Then, a cross-attention structure is employed to fuse the features of the multi-source data with the ERA5 maps. 
This enables MODS to generate ERA5 maps that are more in line with real-world meteorological condition during the sampling process.
In the sampling process, LR ERA5 maps and station observations are utilized as multi-guidance to dynamically control the loss of details and the differences in meteorological states in the generated results.
Experiments have verified that the setting of using multi-source data as conditions in MODS can generate high-quality HR ERA5 maps with faithful details and more realistic meteorological variables. 

\newpage

\bibliographystyle{unsrtnat}
\bibliography{main} 

\newpage

\end{document}